\documentclass[sigconf]{acmart}

\usepackage[utf8]{inputenc}

\usepackage{xcolor}
\newcommand{\new}[1]{\textcolor{black}{#1}}

\AtBeginDocument{%
  }

\copyrightyear{2025} 
\acmYear{2025} 
\setcopyright{rightsretained}
\acmConference[CHI EA '25]{Extended Abstracts of the CHI Conference on Human Factors in Computing Systems}{April 26-May 1, 2025}{Yokohama, Japan}
\acmBooktitle{Extended Abstracts of the CHI Conference on Human Factors in Computing Systems (CHI EA '25), April 26-May 1, 2025, Yokohama, Japan}\acmDOI{10.1145/3706599.3716235}
\acmISBN{979-8-4007-1395-8/2025/04}

\begin{document}

\title{The Cloud Weaving Model for AI development}

\author{Darcy Kim}

\affiliation{
\institution{Watson Foundation}
  \city{New York City, NY}
  \country{USA}}

\affiliation{
   \institution{Socially Intelligent Artificial Systems Group\\
   University of Amsterdam}
   \city{Amsterdam}
   \country{the Netherlands}}
   \email{darcy.kim@wellesley.edu}

\author{Aida Kalender}
\affiliation{
  \institution{
  Socially Intelligent Artificial Systems Group \\ Informatics Institute \\ 
  University of Amsterdam}
  \city{Amsterdam}
  \country{the Netherlands}}
\email{a.kalender@uva.nl}

\author{Sennay Ghebreab}
\affiliation{
  \institution{Socially Intelligent Artificial Systems Group \\ 
Informatics Institute \\ 
  University of Amsterdam}
  \city{Amsterdam}
  \country{the Netherlands}}
\email{s.ghebreab@uva.nl}

\author{Giovanni Sileno}
\affiliation{
  \institution{Socially Intelligent Artificial Systems Group\\ 
 Informatics Institute \\ 
  University of Amsterdam}
  \city{Amsterdam}
  \country{the Netherlands}}
  \email{g.sileno@uva.nl}

\renewcommand{\shortauthors}{Kim et al.}

\begin{abstract}

While analysing challenges in \new{pilot projects developing} AI \new{with marginalized} communities, we \new{found it} difficult 
to express 
them within commonly used paradigms. We therefore constructed an alternative conceptual framework to ground AI development in the social fabric --- the \textit{Cloud Weaving Model} --- inspired \new{(amongst others)} by indigenous \new{knowledge,} 
motifs from nature, and \new{Eastern} traditions. This paper introduces and elaborates on the fundamental elements of the model (clouds, spiders, threads, spiderwebs, and weather) and their interpretation in \new{an} AI context. The framework is then applied to comprehend 
patterns observed in co-creation pilots approaching \new{marginalized} communities, \new{ highlighting} neglected yet relevant dimensions for responsible AI development.

\end{abstract}

\ccsdesc[500]{Social and professional topics~Cultural characteristics}
\ccsdesc[500]{Applied computing~Law, social and behavioral sciences}
\ccsdesc[500]{Human-centered computing~Collaborative and social computing theory, concepts and paradigms}
\ccsdesc[500]{Computing methodologies~Artificial intelligence}

\keywords{AI development, marginalized communities, 
AI democratization, participatory design, community-centred design} 

\received{05 December 2024}

\maketitle

\newpage 

\begin{flushright}
\textit{The Master's Tools Will Never Dismantle the Master's House}.\\  Audre Lorde \cite{audre}
\end{flushright}

\section{Introduction}
The present article emerged as a result of discussions surrounding CommuniCity, 
a three-year project (\new{2022--2025}) funded by Horizon Europe, 
\new{aiming to create} 100 tech pilots for \new{marginalized} communities in various European cities, \new{ with} 
an explicit focus on co-creation. 
The \new{short-term,} temporary nature of the pilots (which last six months), the unbalanced understanding of the concept of co-creation among all stakeholders, and the scarcity of personal narratives about marginalization in CommuniCity\new{'s} communication\new{s} prompted us to spend considerable time discussing the complexities of marginalization on the one hand, and the limitations of contemporary AI solutions' design/development practices on the other.\footnote{During the first two years of the project, UvA researchers conducted dedicated observations to the preparations for \new{two} 
rounds of open calls, gaining insights into the process of creating tech pilots for marginalized communities within CommuniCity. This has been complemented with analysis of the publicly available data found on the project's official website and other relevant documents.} Through the CommuniCity project, we noticed a fundamental friction between the goal of CommuniCity (i.e. serving marginalized communities) and \new{the framework\new{(s)} underpinning} its practical \new{processes}. 
\new{Common} \new{contemporary} practices \new{in the business/tech sector} 
\new{stand often in} opposition to social justice \new{principles.}  Starting from the belief that continuous engagement with marginalized communities is essential---not only during the short-term piloting of tech solutions but also as a permanent practice---
to create trust and sustainable technological solutions that respect the experiences of marginalization, \new{we acknowledged the need for alternative conceptual frameworks, providing complementary views on these matters, and more accessible vocabularies to express the resulting interpretations. In this endeavour,} we found inspiration \new{in various sources, including} nature and teachings expressed by contemporary Indigenous researchers and in \new{Eastern} traditions.\footnote{\new{The authors recognize that what we refer here as ``Indigenous'', ``Eastern''  
traditions are varied, hyper-
localized, and thus evade a monolithic interpretation or canon. It is impossible to wholly represent these diverse traditions in a single paper. This paper instead seeks to show in this academic space that these (marginalized) ideas are relevant, and to inspire further research based on “othered” knowledge systems. 
These few sources and interpretations do not encompass the living reality of these rich traditions, but this acknowledgement should not discourage initial engagement.}}

The method we apply in this paper \new{draws from reflection and introspection \cite{Kaleidoscopy}} and can be connected to radical imagination \cite{radicalimagination}, described as ``{the ability to envision the world, life, and social institutions not as they are, but as they might otherwise be. This process is a collective endeavor through shared experiences, languages, stories, ideas, art, and theory. By collaborating with those around us, we create multiple, overlapping, contradictory, and coexistent imaginary landscapes—horizons of common possibility and shared understanding. These shared landscapes both shape and are shaped by the imaginations and actions of the individuals who partake in them}'' \cite{radicalimagination}. 
To articulate a “multiplicity of Indigenous knowledge systems and technological practices”, the authors of \textit{Indigenous Protocol and AI} use a variety of “heterogeneous texts that range from design guidelines to scholarly essays to artworks to descriptions of technology prototypes to poetry.” Their embracing of personal stories (autoethnographies) and non-technical forms of language in an “unruly format” evoke embodied knowings that lend itself to discussions where lived experience contextualizes agency, methodology, and conclusions. Further, their methodology “encourages discussion that embraces that multiplicity,” by honoring the personal nuances of embodied knowledge that are often discarded in \new{W}estern notions of truth and reason \cite{IP4AI}. 

The process of writing this paper \new{was therefore} grounded in constant dialogue and empathy for the lived personal experiences of fellow researchers. It represents an exercise in radical imagination for a different model of \new{conceiving and} implementing new technologies---one that draws inspiration from 
experiences of personal marginalization, \new{and from various sources, including non-Western traditions and  other non-mainstream conceptualizations}.\footnote{For a matter of presentational effectiveness, and balance amongst the conceptual, observational/interpretational, and personal elements, the present text ended up containing only traces of our personal experiences, although they remain expressed in symbolic forms. This struggle in itself triggered additional how/why questions, left aside for the scope of the paper.}

Our arguments unfold from highlighting the tensions between (the development of) AI technologies and inherent aspects of human nature. Attempting to overcome the root of these tensions, we introduce the \textit{Cloud Weaving Model} and elaborate on its fundamental elements and its interpretation. This abstract model is applied to analyse selected pilots from the  CommuniCity project. 
\new{The proposed interpretive metaphor is designed with the purpose of centering marginalized communities when analysing concrete AI development pilots. Therefore, the resulting 
framework does not aim to be a practical model, but primarily a tool to facilitate reflection on  overlooked social aspects influencing practical AI development. Yet, a}s a result of this analysis, the paper \new{highlights} 
shortcomings of piloting models taken from software engineering \new{standard practices} when applied to the social domain,  
invit\new{ing} for 
a radical rethink of AI technology development \new{to} enable more inclusive approaches. 

\section{Background}

The modern AI paradigm is grounded in “the anthropocentric cultural idea that humans are active agents exerting control over their environments”, which is a symptom of a human-ecological relationship based on resource and knowledge extraction \cite{chi-ecologicalAI}. Furthermore, “Given the long history of technological advances being used against Indigenous people,” \cite{IP4AI}, it is imperative that we engage with this latest technological paradigm shift to actively confront this harmful (colonial) normative structure. Our sustainable and equitable future depends on it \cite{IP4AI}.

The word \textit{techne} relates to use of a tool, some type of intermediary, to gain leverage when crafting. Amongst others, Plotinus talks about how \textit{humans’ love of techne drew them away from being, where we serve an image of nature and not true nature itself} \cite{plotinus}. Gary Snyder wrote a beautiful poem “Ax Handles” about the making of ax handles as extensions of the human body \cite{ax-poem}. This would be the expression of "craft" as distinct from "production", or the "particular" as distinct from the "universal". The ax handle itself was created for a particular hand and arm and body and personality in mind. It was not \new{produced} for an averaged/universal/scalable body that leaves no space for the \new{intrinsic qualities of being} like current technologies \new{do}. BigTech makes phones, laptops, and other products for an "averaged body" which is imagined and separate from the actual body. This observation covers not only statistical-based digital systems as recommendation systems, but also concepts as Industry 4.0 (introduced to pinpoint to a “4th industrial revolution”), where personalization/customization-oriented production systems cater to individuals’ needs \cite{I4a,I4b}. The instances of products or services these systems provide are still based on the image of the user that is primarily formed through quantified data captured by companies about all their customers, sales, operations, and transactions, but not the person\new{s} themse\new{lves}.

Technology amplifies the inertia of the space it is implemented in. It is not neutral, since its creation is dependent on underlying assumptions \cite{cooley}. It is a tool that does what we tell it to do, not necessarily what we want it to do. It is a tool that does what we tell it to do even divorced from the context of being and also in the service of desires of which we are unconscious. Shoshana Zuboff argues in \textit{The Age of Surveillance Capitalism} that the inertia of BigTech and the modern AI Paradigm is essentially violent capitalist-colonial doctrine \cite{survCap}. In our unconscious ignorance, we perpetuate the existing anthropocentric, colonial paradigm.  Evolving from industrial capitalism, which exploits and controls nature, surveillance capitalism exploits and controls human nature to create a totalitarian reality where everyone behaves as this “averaged body”. Other researchers, like Timnit Gebru, do extensive work on how AI algorithms discriminate based on classes like race and gender, since the Big Data used to train these AIs reflects the oppressive, discriminatory nature of society. When these AIs are then given agency to make decisions, they perpetuate existing misogynistic, racist, and classist systems of oppression. \cite{tg1,tg2} The current AI paradigm becomes an assault on human autonomy and democracy. 

In this context, efforts like \textit{Indigenous Protocol and AI} further explain that Western rationalist epistemologies are not enough to break free of this inertia: “If we insist on thinking about these systems only through a Western techno-utilitarian lens […] \new{a}t best, we risk burdening them with the prejudices and biases that we ourselves still retain. At worst, we risk creating relationships with them that are akin to that of a master and slave.” \cite{IP4AI} Proceeding along this program, we must seek bottom-up alternatives to build AI for social good that actively confronts this inertia using knowledge systems beyond the \new{Western mainstream conceptualizations}.



There is history of community-centered scientific and technological pursuits enacting positive social change. Ruha Benjamin, for instance, articulates how collective, community-centered, small scale actions have consistently created positive social change for these {marginalized} people in her book \textit{Viral Justice} \cite{viraljustice}. She counters the assumption that scalability is the most effective way to serve {marginalized} peoples and provides examples of successful alternatives for social justice and change. Ali Alkhatib also explores, “When systems and data disempower other stakeholders, we see absurd and in some cases terrible outcomes. But in contexts [...] where stakeholders are regarded as continuing participants in the ongoing construction of the world that these systems create,” \cite{absurd} the possible harms of these outcomes are mitigated. Cooley explains, “When we design technological systems, we are in fact designing sets of social relationships, and as we question those social relationships and attempt to design systems differently, we are then beginning to challenge in a political way, power structures in society” \cite{cooley}. Agreeing with these perspectives, we ground our search on community-empowering, small-scale approaches, to directly confront the reality of systemic power structures \new{reflected in AI development}.

\section{The Cloud Weaving model}

Inspired by indigenous \new{research} and epistemologies that “refuse to centre or elevate the human” and “reinforce the notion that, while the developers might assume they are building a product or tool, they are actually building a relationship to which they should attend” \cite{IP4AI}, we seek a framework of concepts and associated practices for which \textbf{a successful AI that serves social good is not a final product, but a practice. } Thus, we introduce in this paper the Cloud Weaving model, rooted in the non-human “wisdom” of spiders and clouds, 
used as a basis for a methodology to develop socially-\new{aware} AI. 

\subsection{Clouds, 
and their relation to AI}
Clouds are deemed respected teachers in Eastern traditions \cite{clouds-crazy,clouds-dogen,clouds-korean}. In the Zen Buddhist tradition, as taught by Suzuki Roshi, Buddhism is a practice where one frees themself of ignorance: “According to Buddhism, the origin of suffering is very deep. Originally, we understand that there is some unconditioned being. But when the unconditioned being is conditioned, [...] When this unconditioned being makes some movement, it is the beginning of ignorance.” \cite{suzuki} As previously articulated, AI perpetuates suffering by upholding the harms of systemic inequality. Although AI is not inherently good or bad (i.e. unconditioned), when AI creators are ignorant of these invisible, oppressive conditions placed on it, AI then ignorantly perpetuates suffering.

To break free of ignorance, Buddhists \textit{meditate}. The practice of meditation is the practice of observing ignorance so practitioners are no longer unconscious bearers of it. Meditation brings people in touch with their being, which evades previously discussed notions of an “average body” estranged from the actual body. Clouds are considered teachers of meditation and wandering, the same instruments that let people free themsel\new{ves} of ignorance. Therefore, in the context of AI, clouds can teach designers to work beyond the doctrines estranged from personal reality. 

This conceptualization of clouds has a direct connection with several traditional practices. \textit{Unsui} are, for instance, “Buddhist monks whose detachment from worldly life has them drifting like a cloud over water.”  The zen poet Ikkyu-Zen assumed the pen-name Crazy Cloud, which is a pun on Unsui. He inspired many people who are sometimes called the Crazy Clouds: “The democratic freedom of the Crazy Clouds reminds us of the unequivocal moral authority of the individual, which, in the last instance, must transcend all formal structures and find its own sovereign expression. No mere rebellious outburst against authority, the attitude of the Crazy Cloud, rooted in the fundamental tenets of Buddhism, and expressed in a life that embodies transience, interdependence, and \textit{shunyata}, proves that every concept, even the most noble, must self-destruct.” \cite{clouds-crazy} 

The crazy cloud way examines how Buddhist notions of “the unequivocal moral authority of the individual,” are not defined by opposition. In the context of AI, creating ethical AI that enacts real social good requires making peace with the oppressive inertia and finding ways to work given the reality of its existence. Similarly, Sultana et al. \new{discuss of} design within \new{oppressive systems such as} patriarchy \new{as making} “a temporary peace with the limitations” in order to “empower within a patriarchal society, rather than against it.” \cite{design-patriarchy} The Crazy Cloud way also advises that any form will eventually perish, so there is no final solution of a “good” AI. Instead, the emergence of ethical AI is always becoming and never-ending.

Buddhist Zen master Dogen uses clouds to examine the nature of ignorance in \textit{Shobo Genzo Zazenshin}. “\textit{We should make a concentrated effort to understand this in detail. Rather than love `the carved dragon'}, [The `carved dragon' (\textit{chôryû}) alludes to the ancient Chinese story of the Duke of She, who loved the image of the dragon but was terrified of the real thing.] \textit{we should go on to love the real dragon. We should learn that both the carved and the real dragons have the ability [to produce] clouds and rain. Do not value what is far away, and do not despise it; become completely familiar with it. Do not despise what is near at hand, and do not value it; become completely familiar with it. Do not take the eyes lightly, and do not give them weight. Do not give weight to the ears, and do not take them lightly. Make your eyes and ears clear and sharp.}” \cite{clouds-dogen} Here, Dogen first advises to “love the real dragon” not just an image of it. Interpreted in the context of AI, this is putting the real person in front of us before the imagined person \new{on which} AI base\new{s} decisions. Sometimes, an AI can have a high percentage accuracy, but fails when implemented in the real world. As critical researchers are increasingly arguing, designers must value real-world impact, over \new{sandboxed} quantified notions of success \cite{racialbias-health}. Dogen advises becoming familiar with what is far away and near at hand without valuing it. Before designing AI in the context of {marginalized} communities, designers must become familiar with both the practices of tech developers and community organizers, without valuing one way over the other. Instead, these stakeholders should first become familiar with one another without making \new{judgments} or jumping to conclusions. Dogen ends this passage by asking us to make our “eyes and ears clear and sharp” without giving them weight. In this sense, he advises people to take their own biases and prejudices into consideration and to sharpen our senses so \new{that} people are not blinded by them.

Korean Seon Buddhist monks also used \textit{baekun}, which means white cloud, as a symbol for respected Buddhist practices.  Gyeonghan referred to himself as \textit{baekun} and wrote poems about \textit{baekun} to convey the realm of "no-mind, no-self." Bou wrote poems of \textit{baekun} as a symbol of Buddhist practices, since white clouds that drift absentmindedly become rain and wet the earth, serving all sentient beings. \cite{clouds-korean}

Sentient beings are beings that feel and suffer, but in the context of AI, they can be considered all beings that are impacted by AI. 
Clouds drifting absent-mindedly without “mind” and “self” are akin to people observing the current AI methodology without acting nor imposing personal beliefs onto it. Essentially, merely observing the harmful ignorance of current AI methodologies, and choosing to not act, is still worthwhile. There is merit in facing the reality of AI's current way, and our own personal beliefs of it, even without acting to change it. Efficiently producing and acting is not the only success. Mutual learning between stakeholders, where they are allowed to wander and be with each other, is in itself valuable. 

As these Buddhist traditions show, clouds are radical wanders that are not defined in opposition (and therefore still by) any form of authority or power structure (\new{nor} of self \new{nor} of mind), but transcend these notions in an effort to end suffering and mitigate harms. In a time when AI is built to serve a specific way of being, clouds may serve as a guide for facing these forces we are ignorant of. They call AI designers to transcend the status-quo and overcome ignorance to create technology that does not perpetuate the suffering of the current oppressive systems.

\subsection{Main elements of the model}
The Cloud Weaving model consists of 5 elements: \textbf{clouds}, \textbf{spiders}, which connect clouds through \textbf{threads} weaving \textbf{spiderwebs}, and the \textbf{weather}.  

\subsubsection*{\textbf{Clouds}}
Clouds are always flowing, never ceasing, never stopping. In our model, clouds represent the peoples, communities, stakeholders, and sentient beings in relation \new{to and} flowing through each other. They include, for instance, the \new{marginalized} peoples centered in the design projects, tech developers, and municipality officials involved in smart city initiatives. They also include the natural resources and energy needed to build and maintain AI, the people whose data is used to train it, and other ecological dimensions, as all these perspectives should be respected in the design process. Like in the  tradition they are inspired from, clouds must  work to observe and free themsel\new{ves} of the harmful \new{ignorance} they carry.

\paragraph{Individuality} It \new{i}s impossible to draw a clear boundary between one cloud and the next since they are always flowing into and out of each other.  Similarly, communities overlap and entwine. It \new{is} impossible to distinguish an individual separate from community, and individual\new{s} uninflected by a relationship beyond themself. Within a cloud (community), it \new{i}s impossible to distinguish an individual. Thus, clouds offer a compelling representation of the stakeholders and beings and communities we must link together with an adequate web. When uplifting \new{marginalized} peoples, we must offer support for the whole community, not just arbitrary individuals within it.

\paragraph{Dynamism and Flow:} Clouds are always moving. Sometimes they curl in on itself and blend with other clouds or separate in skies beyond. They are chaotic systems, meaning \new{that} it is impossible to predict what clouds will look like in the future, since small changes can have big effects. Similarly, society and social relations are always in flux. It is impossible to perfectly predict how clouds will change. To weave a web in the clouds, the spider must always adjust to these constant changes and continue weaving through them. Since these changes are impossible to predict, a socially \new{aware} 
AI must continually respond to new movements, changes, and societal forces (\new{including} politics).

\subsubsection*{\textbf{Spider}}
A spider facilitates clouds to connect, interact, and possibly merge by weaving webs. A spider is the AI that community organizers-designers co-create and represents the AI given agency. The \textit{Indigenous Protocols for AI} describe AI as a being that we should care for and are in ongoing relationship with \cite{IP4AI}. The spider is akin to a baby that the village of clouds are raising.

Spiders have captured the imagination of community activists and tech developers alike for decades. Spiderbots, also called webcrawlers or simply spiders, are computer programs for traversing the internet for indexing,  to discover and to collect information across the world wide web \cite{webcrawlers}. Here, spiders encompass the meticulous nature of searching and discovering across a complex, informationally dense landscape. 

Modern philosophers were also attracted by spiders. Marx draws inspiration from spiders to explore labor as a relation between humans and nature:  “A spider conducts operations that resemble those of a weaver, and a bee puts to shame many an architect in the construction of her cells. But what distinguishes the worst architect from the best of bees is this, that the architect raises his structure in imagination before he erects it in reality.” \cite{marx} Just like webcrawlers, spiders for Marx are operators, lacking imaginary skills necessary for going beyond what they have been determined to do. Yet, their immanence gives them the advantage of being present only where they have been posited onto. In our model, no other worlds for the spider exist, but the one of the clouds upon which the spider is weaving threads (\textit{local presence}). Second, they have no other objective than connecting clouds (\textit{systematic transparency}). From a complementary perspective, Francis Bacon famously referred\footnote{``The men of experiment are like the ant, they only collect and use; the reasoners resemble spiders, who make cobwebs out of their own substance. But the bee takes a middle course: it gathers its material from the flowers of the garden and of the field, but transforms and digests it by a power of its own.'' \cite{bacon}} to spiders in association with metaphysics (dedicated primarily to theoretical/abstract concerns), contrasted to ants in association with empiricism (dedicated primarily to collection of facts to induce empirical laws), and to bees as figurative representatives of the best approach to science (i.e. to devise theories in combination with practice). In our model, the spider is deemed to connect clouds, and it cannot be (part of) a cloud. They are intrinsically different: people and sentient entities, which form clouds, have grounds onto reality that AI cannot (and should not?) ever have. This makes spiders, even if acting as coordinating entities, subject to clouds.

\subsubsection*{\textbf{Thread}}
The thread is the information that the spider/AI uses to connect all stakeholders. If the spider is the AI, it is constantly changing under the weather that inundates it and the clouds that relate to it. From the perspective of the weather, the thread represents how AI decisions are based on its data input. Timnit Gebru explains that data reflects the society it comes from, so an oppressive society can only yield an oppressive AI \cite{tg1,tg2}. However, this thread is not just \new{the} result of quantifiable data, but also the information of clouds that shape the spider: embodied knowings, oral traditions, trust, love, and care. Thus, a thread is forged through “an undetermined articulation of ‘being with’ that involves learning to be affected and attending to difference without reifying that difference once again” \cite{empathy} which has been encouraged in affective partnerships between designers and disabled people. This thread is the raw outcome of the AI: it is both inherently reflective output and also the relationships and trust that create it.

\subsubsection*{\textbf{Spiderweb}}
A complete spiderweb is the “final” goal of socially \new{aware} AI. It is a structural support that serves \new{a} social good being built through AI. (However, due to weather, there is no final web, since it requires constant repair to maintain its sustainability.) Webs change the structure of the clouds (\textit{systemic change}) and `catch' something that falls through in the empty space between them (\textit{individual relief}). 

\textit{The Problem} the AI is solving is represented by the (virtual, not yet connected) center of this web. This center is always shifting as more spokes are added to the web, clouds shift, weather weathers, and the spider continues weaving. Similarly, solving social problems is like hitting a wobbling bullseye since they are always oscillating due to the dynamic nature of society  \cite{dynamic-social}. The practice of prejudice is “mercurial” and “shape-shifting,” \cite{absurd} so adequate responses to it must also be flexible. 

\textit{The Solution} to the problem is solved by completing the web around this center, that is dynamically and incrementally capturing the threads amongst clouds (that is, realizing adequate socio-technical arrangements \cite{AIlaw}) in a way to approach the center. The spider spirals thread into the center without actually ever meeting it. The “solution” to the social problem can never be fully achieved, although we can get closer to the heart of it. 

\subsubsection*{\textbf{Weather}}
The clouds, the spider, and the web are all subjected to weather. This weather encompasses the model and is entirely inescapable. Ruha Benjamin in her book \textit{Viral Justice} \cite{viraljustice} similarly uses weather to describe the systemic power structures \cite{structure} that pervade our entire life, including the technology we build. Weather affects access to resources and the relationship between clouds (stakeholders, people, and the natural world). It also determines whose perspectives are valued, where those of non-humans and earth is often ignored. \new{On similar veins, decades ago, in his ``\textit{The Lost World of the Kalahari}}'' \new{Laurens van der Post aptly framed drastic changes and developments in human life and society as thunderstorms \cite{sennayquote}: ``It seemed that species, society, and individuals behaved more like thunder-clouds than scrubbed, neatly clothed, and well behaved children of reason. Throughout the ages life appeared to build up great invisible changes like clouds and earth of electricity, until suddenly in a sultry hour the spirit moved (\dots) and drums tolled to produce what we call thunder and lightening in the heavens, and chance and change in human society and personality.'' These storms can damage the spiderweb. However, overtime, we can learn to actively fortify the web against these forces. Like a spider, we must persistently repair and restart the making of the web, even when it falls in storms.}

\begin{figure*}[t]
  \includegraphics[width=\textwidth]{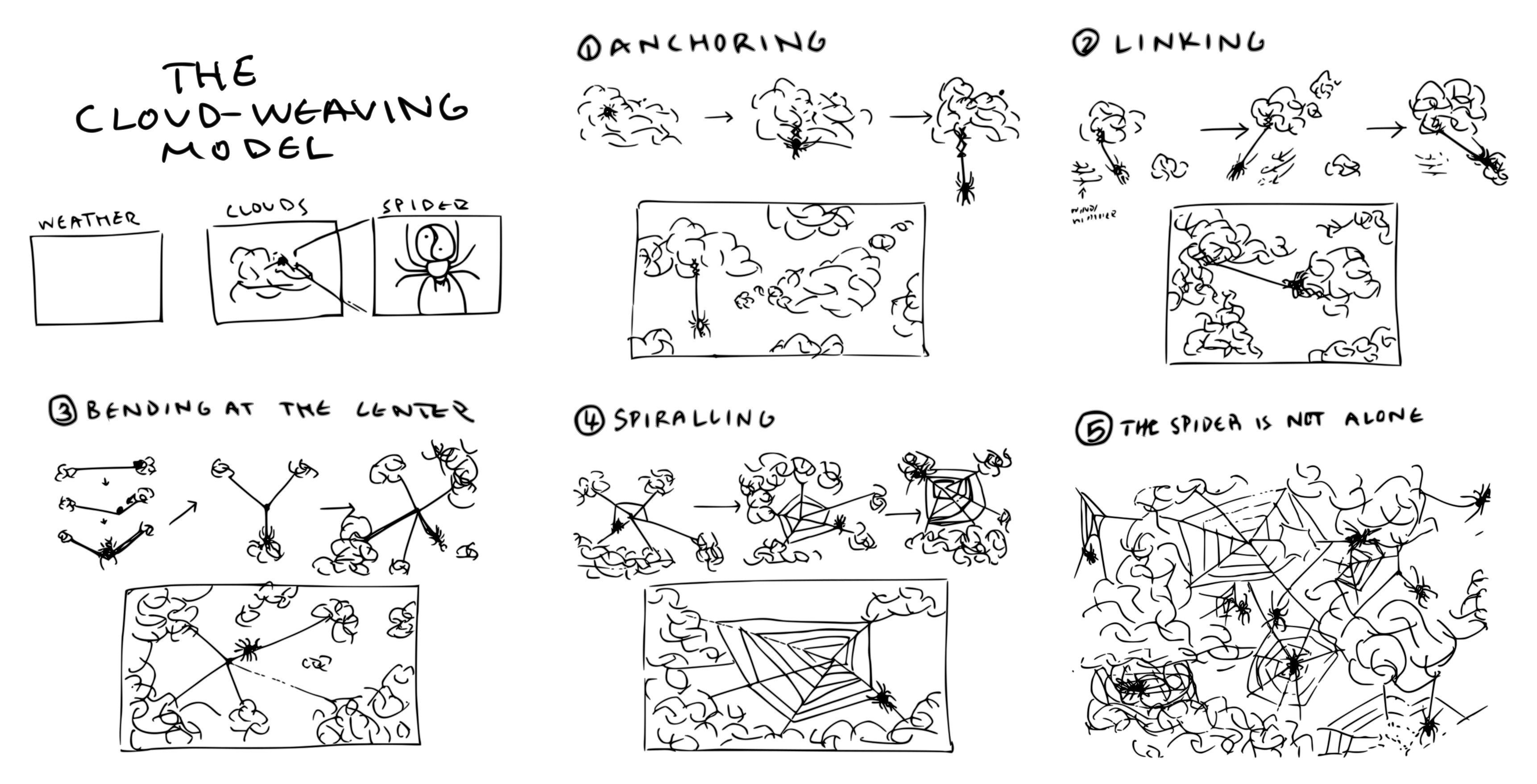}
  \caption{\new{The five processes expressed in the Cloud Weaving Model: \textit{anchoring} (by which the spider connects to a cloud), \textit{linking} (by which the spider connects nearby clouds, although challenged by the weather), \textit{bending at the center} (by which the spider attempts to find a balance across clouds), \textit{spiralling} (by which the spider makes the spiderweb more robust, and approaches the center), \textit{encountering other spiders} (expressing and possibly realizing power relationships)}.}
  \Description[The five processes expressed in the Cloud Weaving Model]{The five processes expressed in the Cloud Weaving Model: \textit{anchoring} (by which the spider connects to a cloud), \textit{linking} (by which the spider connects nearby clouds, although challenged by the weather), \textit{bending at the center} (by which the spider attempts to find a balance across clouds), \textit{spiralling} (by which the spider makes the spiderweb more robust, and approaches the center), \textit{encountering other spiders} (expressing and possibly realizing power relationships).}
  
\end{figure*}

\subsection{How to weave a web}
\new{Based on the prevous five elements, the Cloud Weaving model captures five processes: \textbf{anchoring}, \textbf{linking}, \textbf{bending at the center}, \textbf{spiralling}, \textbf{encountering other spiders}, illustrated in Fig.~1.}

\subsubsection*{\textbf{1. Anchoring.  }}First, the spider must anchor itself in a cloud. However, spiders are not accustomed to weaving amongst clouds, so creating this first anchor will take time. It takes time for a cloud to trust and learn to braid itself into the spider’s thread to keep its connection stable and strong. Overtime, this connection can erode and requires maintenance as clouds change and time passes and weather weathers. Overtime, the spider learns more about how to keep it strong. Building this first strong anchor represents the time and care required to establish that first connection between tech developers and marginalized communities. This is a generation-spanning commitment to maintain. 

Previous HCI research explores effective ways to develop trusting relationships between designers/developers and users/peoples. They center “understanding that emerges from seeing each person as a separate unique centre of value and then responding to them from the special value position that is one’s own. This kind of affective response cannot come from a third-person perspective as it involves aesthetic seeing, not scientific knowing” \cite{Wright2008}. This step requires time for both parties to familiarize with each other and develop a mutual learning from and with each other. They must dedicate time to exploring the existing social fabric and the informal support systems in place. Well-meaning outsiders who neglect this informal social fabric can further disrupt working measures and cause more problems than solutions, especially in smart city contexts \cite{smartcity,ecologiescare}.

\subsubsection*{\textbf{2. Linking or Bridging}} Now that this spider is anchored in a deep relation with the community, it can pull out its own thread and dangle in the wind. The weather swirls it around until it finds another cloud to anchor in. However, weather tends to take this spider to certain clouds. To get to hard-to-reach clouds against the weather (i.e. marginalized communities), the spider must find ways to wander, traverse, crawl hemispheres and altitudes away. It takes active work to go against existing power structures that hide certain communities, and the spider must make an active effort to reach these places. For example, the spider must be actively anti-racist as it forges relationships with different peoples, communities, stakeholders, and sentient beings. 

Like in step 1, it takes time and trust to make a secondary anchor point bridging these clouds together. With every new community integrated in this network, a new anchor must be made. However, every cloud is different, so every anchor point requires time, care, and trust to establish. 

\subsubsection*{\textbf{3. Bending at the center}.} Now that the spider has established this first bridge thread, it walks toward the center and allows the thread to bend under its weight. This point becomes its center, and the spider continues to find more clouds to anchor to building spokes around this center. As more spokes are added, the center shifts under the added tension. Overtime, the position of this central problem becomes more stable. As the spider bridges more communities together in the development of AI, the problem it is seeking to solve becomes clearer. 

As the spider continues to build spokes, the center wobbles. Each time the spider adds a new spoke, it returns to the center and adjusts the tensions of each thread. After each new connection with an additional community, the spider must also actively reflect on the nature of the problem and adjust accordingly. Over time, as new spokes are added and tensions are continually adjusted, this center stabilizes.

\subsubsection*{\textbf{4. Spiralling.}} Now that the spider has stabilized the center, it can start making a spiral towards it. \textit{Spirals are for hunting, straight lines are for destinations.} There is no destination in social transformation, since it is an ever-unfolding process. Instead, we are persistently hunting for ways to build and maintain support systems. 

Now that AI has a clearer understanding of the problem, it can start finishing its solution to it. It begins to spiral into the center. It is not a perfect counterclockwise spiral. Maybe there is more space on one side since the center is offset. The spider turns back on itself to make up for this space. Similarly, the spiral to the center is not unidirectional. Sometimes, the spider has to turn back on itself and re-evaluate the choices it has made.

\subsubsection*{\textbf{5. The Spider is not alone.}} In a complex social system, we can expect many spiders to be grown, each building and rebuilding their own webs. It is inevitable that spiders run into each other. Furthermore, since AI models are sometimes built on top of other AI models, spiderwebs can be inside another spiderweb. Chang et al. observed that, in nature, ``aggressive spiders made fewer directional changes before completing tasks, regardless of the task's difficulty. However, decision accuracy was jointly determined by both aggressiveness and task difficulty. Aggressive spiders made more accurate decisions in the simple task, while docile spiders made more accurate decisions in the difficult task.'' \cite{spiderfight} Aggressive spiders are less inclined to flexibility and do more poorly in difficult tasks. Similarly, we should guide spiders to remain flexible and soft in the engagement with one another. However, conflict is inevitable, in both the animal kingdom and the human domain. Conflict has also the potential to bring people closer together, and can possibly strengthen these webs of webs. It is not something to be avoided, but something to address head on.

Technically, a local AI (a spider whose web is not nested) should also protect local authenticity from the attack of globalizing \new{forces}. \new{Spiders do not act hierarchically---each spider nurtures different ``centers'' within the web of webs. The collective cultivation of all these centers working together transforms the AI paradigm from within, rather than enforcing a new doctrine upon all creations.}

\subsection{Differences from physical webs}
\subsubsection*{\textbf{Sustainability.}}  A spider is never able to meet the center (just as clouds and weather keep changing). It is doomed to constantly circle it without ever reaching it. AI solutions to social problems are never done, they have to be seen as practices. It is impossible to complete a web. Weather, an unceasing force, tears the model to disrepair and shifts the clouds that the web is anchored to. Over time, the threads become worn and weak. The web requires constant maintenance and repair, actions which further take out resources from its completion. 
AI integrated in social settings needs to constantly reflect on how it affects the communities it bridges, add new communities to gain a better understanding of the central problem, and update itself to remain standing amidst all this inevitable change.

\subsubsection*{\textbf{(No) Catching Flies.} }The spider and spiderweb itself is the solution. Spiderwebs are designed to catch flies to feed the spider. When the spider instead weaves the clouds, it is used for something it was not initially meant to do. There are not many flies high in the sky. To sustain itself, the spider must work more slowly and intentionally under these conditions,
not to waste precious energy. It might even need to rely on friends to bring it food when there are not enough resources here. 

The modern AI paradigm was forged under the stress of economic survival. Under this doctrine, it is best suited for making profit. 
When used to serve marginalized peoples who have less resources, it must be made more intentionally and carefully to avoid burnout.

\section{Application: The CommuniCity project}
Technological developments in the city context are transforming private and public organizations, impacting service provision in all aspects. Several efforts have been started to promote a “democratisation” of these processes of innovation \cite{Toward}, but generally they are meant to involve resourceful, engaged, tech-savvy parts of the population. Despite being commendable, this is not sufficient to fulfill the promise of fairly distributing the opportunities of technological transformation towards citizens and the common good, even more so when we consider marginalized communities. New innovation methods and approaches are urgently needed.

The CommuniCity project, currently ongoing, originated from this requirement. Its goal is to run 100 experiments through pilots of technological co-creation targeting marginalized communities across Europe. The project has undergone three iterations, and the initial pilots have revealed the significant challenges of technological transformation, particularly with regard to the inclusion of marginalized communities. Involved stakeholders, ranging from civil servants in municipal administrations to officials in innovation agencies, tech developers, and representatives of marginalized communities, have demonstrated the difficulty and complexity of their tasks. 
  
While analysing the CommuniCity pilots \cite{ComCityMain}, we identified some relevant patterns that can be revisited through the Cloud Weaving model. These patterns were selected here not to be exhaustive in describing the project, but rather to exemplify the value of changing paradigm while self-reflecting over a such complex applied domain.  

\subsection{Pattern 1: Establishing Enduring Threads}  

One of the observed pilots aims to improve the socio-emotional skills of young people in vulnerable communities by means of virtual reality (VR) games. 
Existing VR games are used to explore new possibilities to enhance communication, stress management, and empathy skills.
For doing this, youth workers are trained to steward the game and create a manual to share knowledge for potential new youth workers beyond the pilot. 

In this pilot, the \textbf{weather} encompasses the social, political, and economical factors that young people’s mental health (and empowerment in the face of) relies on. Jiddu Krishnamurti famously said, “It is no measure of health to be well adjusted to a profoundly sick society.” \cite{sick} In this way, youth mental health depends on more than an individual’s brain chemistry, but also their environment and societal expectations.

The \textbf{clouds} include tech developers, young people, and youth workers who were part of the co-creation process. It also includes the companies and policy-makers in charge of challenge formation. However, although they were not considered in this piloting process, the clouds could also encompass the families, educators, friends, mentors, sports teams, and pets in these young people's lives. They could include the beings they engage with on social-media, which also has a huge effect on mental health \cite{mentalhealth}.  To draw an even wider scope, the clouds encompass the natural resources (and the people who extract them) used to make VR and the energy/carbon emissions used to make and sustain them. 

The \textbf{spider} is the making of the VR application itself (the AI development practice in place). The \textbf{spiderweb} that the spider is weaving includes the relationships forged through the making of the VR and the conversations inspired by the using of it. It encompasses the social integration of VR where youth workers directly facilitate conversation with youths surrounding their experience. It can also include new relationships forged through the playing of VR games and also the strengthening of existing relationships through the practice of VR.
The pilot also worked to create lasting social support through the training of youth workers and by making a manual for potential youth workers in other cities. In doing so, their impact goes beyond their tech product by directly investing in community’s future through social, non-tech-centered means of support.

A possible \textbf{breaking of the web} might elucidate in post-simula-tion reflections as an avenue to explore accountability and uncover where the VR does not work. Girls’ experiences can vary significantly from boys’ experiences since children are raised gendered, which includes differing emotional responses and behavioral values. This changing \textbf{weather} of gender expectation inflicts the experiences of youths: what it means to be a good `girl' vs a good `boy' changes with time. Repairing this web by incorporating reflection, which yields new information to continually update and improve on the VR, can make  it sustainable for future generations.

Another \textbf{hole in this web} is funding caused by the \textbf{weather} of economic scarcity.  Currently, CommuniCity provides no exit strategy or support after the six-month developing process, so webs fall to disrepair when resources to continue their maintenance evaporate. Additionally, there are still few grants available to further research or promote the use of social technology.

Although this pilot operated under a tech/business centric framework, their commitments to co-creations reflect efforts for communi-ty-centered, long-term partnerships that the Cloud Weaving model recommends. Cloud Weaving also offers a framework to explore the social implications of their technology beyond the completion of their product, inviting additional imagination and reflection necessary to ongoing social support.

\subsection{Pattern 2: Apps and Platforms Weave Fragile Webs}
Examples of \textbf{spiders} in the observed pilots include custom community engagement platforms. One promotes girls' participation in sports, while others address issues such as elderly loneliness, youth mental health, community exchange, and urban accessibility. In these scenarios, the corresponding \textbf{spiderwebs} are the relationships and interactions that these platforms facilitate, along with the events and connections they generate.  For example, for the girls’ participation in sports pilot, the spiderweb includes the sports games and teams, along with the relationships between coaches, parents, and young athletes. The \textbf{clouds} of these social media platforms are all relevant stakeholders beyond the users and platform creators. For the girls’ participation in sports pilot, it also includes the young girls, the coaches, their parents, and also the stewards of the athletic spaces they use. For another pilot, a platform that seeks to decrease elderly loneliness, the clouds include elderly people, their friends and family and pets, caretakers, doctors, and also government support for elderly people. 

The \textbf{weather} surrounding these platforms is systemic in nature.  However, social media platforms primarily enable individual empowerment, limiting their effectiveness in addressing systemic forms of oppression. These platforms rely on individual participation without providing systemic support or incentives for engagement. For example, the girls' sports platform assumes young girls want to play sports but lack the means to organize, overlooking structural barriers such as unequal infrastructure and access between boys' and girls' sports. Additionally, platform-based solutions often ignore existing online community-building mechanisms, such as popular social media platforms, and fail to examine why these avenues are underutilized.

\subsection{Pattern 3: Prioritizing Sustenance Over Clouds}
One of the pilots aimed to support facilities for refugees through a mobile app.  In this pilot, the \textbf{spider} was developed as a mobile app that provides translation services for refugees. The \textbf{spiderweb} would be the systems of social support that the app opens for refugees who often face administrative challenges. However, this spiderweb can easily fail to seriously incorporate all relevant \textbf{clouds}, most notably refugees themselves. Several dimensions explain the distance (and thus the difficulties to reduce the gap) between refugees and app developers: language and communication barriers, cultural sensitivity, technological and practical accessibility, trauma, psychological and trust barriers. 

Thus, the \textbf{center} of the web, representing the problem, is uninflected by the very people the app hopes to serve. Their lived experiences and the reality of their problems are unacknowledged in the building of the web if adequate resources are not spent to reduce the gap. In part, this is due to the \textbf{weather}, which encompasses the need to generate revenue to sustain the project. This requirement pragmatically takes precedence over the acts of being with refugees and slowly building trust and mutual learning, to the detriment of the effectiveness of the overall endeavor. 

\subsection{Pattern 4: Spiderwebs Breaking and Rebuilding}
Throughout the piloting process, some spiderwebs were broken by weather. The pilot \new{developers who built} a platform based on interactive videos \new{to} understand and improve \new{the} digital skills of unemployed people pointed to a `knowledge gap' — experienced while setting up the platform — as a main struggle in co-creation. Existing social, economic, and political factors (i.e. the weather) prevent tech developers and these unemployed peoples from receiving the same kind of education, and thus create communication struggles. 
Another pilot also cited a language barrier as a hindrance in co-creation. The \textbf{weather} that leads people to knowing different languages can create communication struggles that tear through the connections of the spiderweb. In another pilot, it was observed that maintaining the motivation and concentration levels of homeless people to use digital devices was very difficult, since the stresses related to homelessness (the weather) drain energy and concentration capacity.

Nevertheless, developers of one pilot provided an example of rebuilding a spiderweb which had been destroyed as soon as the misalignment of expectations with the target group became manifest. The original plan was completely discarded, and development proceeded towards what the target group actually wanted and needed. In other words, the original spiderweb was destroyed after learning of the actual needs of underprivileged peoples. In this process, the tech developers involved declared that they have deepened their understanding of requirements for designing for marginalized groups. This strategically strengthens the spiderweb to which they contribute in the face of weather.

\subsection{Pattern 5: No Web Possible}
One of the pilots reached the conclusion that a technical solution was insufficient for the problem. Specifically, the linguistic support that the target solution was asked to provide was too critical in the face of existing linguistic gaps. In this case, the spiderweb could not be formed, as the weather conditions were too extreme for the spider alone to sustain. \new{Sometimes, the most advanced technology is not the best solution, and in this sense, non-technical solutions must also be explored alongside technological solutions.} 

\section{Perspectives}

Observing stakeholders involved in pilots developing AI with margi-nalised communities, we noticed several challenges and tried to comprehend their underlying structure. In this paper, we revisited them through the metaphor of weaving a spider's web among the clouds. Through this model, we can argue that the weaving is not in vain; it never stops, even though weather disturbances usually tear the webs apart. The perseverance of the spider, rather than the short-lived, one-off piloting, should be the motivation for all involved in development of socially-\new{aware} 
AI.


\new{In general, metaphors present translation challenges due to their linguistic and cultural complexities, leaving the translation process intrinsically open to diverse cultural interpretations. However, people not only talk metaphorically, they also think and feel metaphorically. As metaphors 
have their basis in bodily interactions with the physical world \cite{0fa64ddf-1eab-36c6-af38-c71ec3c3d4f7}, they  have the potential to alter how we think and feel about various topics. \cite{7bbf9e97-a9f2-3051-b9d5-6da2f6bfb039}}. 

The metaphor expressed by the Cloud Weaving model invites practitioners to envision ethical and responsible AI development as a relational process, rather than an endpoint: AI as a practice, rather than a product.
\new{Yet, this model originated  being grounded in a very concrete localised perspective, i.e. the case studies of the CommuniCity project.  
Although numerous perspectives from various stakeholders in different European cities were entangled in the CommuniCity discussions, aiming for completeness (e.g. by enriching the metaphor with additional perspectives) was not our core target. The model does not seek to universalize. It does not furnish a one-size-fits-all solution. Instead, it seeks to bring to the foreground the conceptual limitations of contemporary tech-centered epistemologies, and potentially assists in reimagining evaluative frameworks for projects like CommuniCity to better achieve their goals. The model at this stage serves primarily as a interpretative tool, meant for critical reflection. 
Whether and how it can be transformed into practical guidelines and effective recommendations for  practitioners, designers/developers, or policymakers is left to future research.} 
As a first step, we will explore 
how the Cloud Weaving model maps to the guidelines for ethical AI design with marginalized communities, currently being developed within the CommuniCity project (e.g. \cite{Toward}.)



A key insight expressed by the Cloud Weaving model is that AI integrated in social settings needs to constantly reflect on how it affects the communities it bridges, add new communities to gain a better understanding of the central problem, and update itself to remain standing amidst all these inevitable changes. 
\new{If we measure the scale of society as the number of roles it exhibits, adding AI (designer/spider- weaving the clouds) corresponds to an increase of scale. As observed in anthropology, the scale sets limits to the scope of options for action, but simultaneously it is the product of action \cite{da4f9b10-a4e1-3652-a285-9cc201c00f40}. This view converges with the image of socio-technical innovation suggested by the Cloud Weaving Model, where more and more spiders weave simultaneusly, creating a continuous movement on a web of webs.}

In contrast, contemporary AI solutions (and even more pilots, for the lower resources available) are developed primarily following software engineering and project management practices, oftentimes informed by the Agile methodology (or at least terminology). When piloting is applied systematically, as in contexts like CommuniCity and similar public efforts, it is regimented into a rigorous process by following predefined steps and procedures that improve the effectiveness/efficiency tradeoff and give priority to outputs. But what is an output, when AI is seen as a practice? \new{Although merely a reflective tool, the Cloud Weaving model may already encounter resistance from stakeholders in AI development, due to the rigidity of the current piloting format.
\footnote{\new{A similar situation was observed during the CommuniCity project when the proposal of a simpler card set meant to facilitate reflection encountered some resistance and sparked discussion amongst partners due to differing terminology.}} However, the Cloud Weaving model’s use of familiar and straightforward concepts could hopefully encourage critical thinking.}

\new{Beyond CommuniCity, the proposed} descriptive framework provides perspectives for alternative paths for AI development. \new{It opens conversations about} aspects which are today generally overlooked and consequently under-resourced. For instance, maintaining documentation of interactions with marginalized communities (even if not successful in the short-term) could inspire future developers and researchers in the field. Communicating weaving relationships with these communities would enhance trust and reduce distances. This could be realized by several means: by showcasing stories of genuine connections between tech developers and marginalized communities, by setting up educational projects in hard-to-reach areas, by providing scholarships for AI studies for children from marginalized communities, and by giving space to personal lived experiences about marginalization but also of (potential) re-empowerment. These narratives would be in stark contrast to --- and provide a more socially sustainable vision than --- the hype surrounding AI, as omnipotent Other with respect to people, putting a much greater emphasis on care, support, and humanity.


\begin{acks}
This work has been supported by the European Union’s Horizon Europe research and innovation programme under grant agreement No 101070325 (Innovative Solutions Responding to the Needs of Cities \& Communities [CommuniCity], 2022).

This work has also been supported by the Watson Foundation funding the Thomas J. Watson Fellowship, a one-year grant for purposeful, independent exploration outside the United States, which allows Darcy Kim to partner with the Civic AI Lab and the Socially Intelligent Artificial Systems Group at the University of Amsterdam.

Thank you to John Bailes and Ehyun Kim, knowledge bearers of Zen Buddhist and Korean traditions and ways of beings, who provided guidance under the teachings of clouds. 

\end{acks}

\bibliographystyle{ACM-Reference-Format}
\bibliography{sample-base}


\end{document}